\def\mathcal{\cal}
\documentstyle[prl,aps,multicol,eqsecnum]{revtex}
\input psfig

\newcommand{\bleq}{\ifpreprintsty
                   \else
                   \end{multicols}\vspace*{-3.5ex}{\tiny
                   \noindent\begin{tabular}[t]{c|}
                   \parbox{0.493\hsize}{~} \\ \hline \end{tabular}}
                   \fi}
\newcommand{\eleq}{\ifpreprintsty
                   \else
                   {\tiny\hspace*{\fill}\begin{tabular}[t]{|c}\hline
                    \parbox{0.49\hsize}{~} \\
                    \end{tabular}}\vspace*{-2.5ex}\begin{multicols}{2}
                    \fi}
\newcommand{\bcols}{\ifpreprintsty\else\begin{multicols}{2}\fi}
\newcommand{\ecols}{\ifpreprintsty\else\end{multicols}\fi}

\def \be{\begin{equation}}
\def \ee{\end{equation}}

\def \ve{\varepsilon}

\def \HH{{\mathcal H}}
\def \GG{{\mathcal G}}
\def \SS{{\mathcal S}}
\def \tr{\text{tr }}
\def \Seh{\Sigma_{eh}}
\def \She{\Sigma_{he}}
\def \See{\Sigma_{ee}}
\def \Shh{\Sigma_{hh}}
\def \GN{Z_N}
\def \GS{Z_S}
\def \IGap{\widetilde{\Delta}}
\def \ETild{\widetilde{E}}
\begin{document}

\bibliographystyle{simpl1}

\title{Andreev Conductance of Chaotic and Integrable Quantum Dots}

\author{A. A. Clerk, P. W. Brouwer, and V. Ambegaokar}
\address{Laboratory of Atomic and Solid State Physics,
Cornell University, Ithaca NY 14853, USA} \maketitle
\centerline{April 19, 2000}
\begin{abstract}

We examine the voltage $V$ and magnetic field $B$ dependent
Andreev conductance of a chaotic quantum dot coupled via point
contacts to a normal metal and a superconductor. In the case where
the contact to the superconductor dominates, we find that the
conductance is consistent with the dot itself behaving as a
superconductor-- it appears as though Andreev reflections are
occurring locally at the interface between the normal lead and the
dot. This is contrasted against the behaviour of an integrable
dot, where for a similar strong coupling to the superconductor, no
such effect is seen. The voltage dependence of the Andreev
conductance thus provides an extremely pronounced quantum
signature of the nature of the dot's classical dynamics. For the
chaotic dot, we also study non-monotonic re-entrance effects which
occur in both $V$ and $B$.

\end{abstract}

\bcols

\section{Introduction}

Though the actual process of Andreev reflection is simple to
describe -- an electron in a normal metal incident on a
superconductor is reflected back as a hole \cite{ANDREEV} --, it
serves as the fundamental basis for some of the most striking
effects known in mesoscopic physics \cite{BeenRMT}.  In
particular, Andreev reflection may be viewed as the phenomenon
underlying the proximity effect, in which a superconductor is able
to strongly influence the properties of a nearby normal metal.
Andreev reflection is also the process responsible for the unique
conductance properties of normal--superconducting (NS) junctions,
such as the conductance enhancement observed in NS point contacts.

Attention has recently turned to so-called ``Andreev billiards'' as
ideal systems in which to study the proximity effect
\cite{GOLDBART,FRAHM,MELSEN,ALTLAND,LODDER,BROUWERJJ,SCHOM,RICHTER}. Such
systems consist of an isolated normal metal region, small enough
that electrons remain phase coherent within it (i.e. a quantum
dot), coupled weakly through a point contact to a superconducting
electrode. Calculations of the density of states in such
structures have shown, rather remarkably, that the proximity
effect is sensitive to the nature of the classical dynamics-- a
gap of size $\IGap$ in the
spectrum is present in the case of a chaotic billiard,
whereas in the integrable case the density of states vanishes
linearly at the Fermi energy \cite{MELSEN,LODDER}.

In the present work we also focus on the Andreev billiard, but now
add a second point contact leading to a normal electrode and
investigate the Andreev conductance of the resulting structure.
The Andreev conductance is the differential conductance $dI/dV$ at
voltages smaller than the superconducting gap in the S electrode,
where Andreev reflection at the interface between the normal metal
dot and the superconductor is the only current carrying mechanism.
Unlike previous studies \cite{ARGAMAN,BROUWERJMP}, we calculate the full
voltage ($V$) and magnetic field ($B$) dependence of the
conductance.

Two questions are of particular interest in this study.  First,
does the sensitivity of the proximity effect to chaotic versus
integrable dynamics, as seen in the density of states, also
manifest itself in the Andreev conductance?  We find that indeed
it does. In the case where the contact to the superconducting
electrode is much wider than that to the normal electrode, the $V$
and $B$ dependent conductance of a chaotic dot is consistent with
assuming the dot has itself become superconducting.  The
conductance is doubled with respect to the normal state, and
remains bias-independent until $eV$ reaches the induced gap
$\IGap$ in the Andreev billiard.  This is in sharp contrast to
what is found in exact calculations for integrable dots.  For both
rectangular and circular dot geometries, the conductance drops off
linearly with voltage without any plateau near $V=0$.   We argue
that this difference is ubiquitous for integrable versus chaotic
systems.

The second motivation for the present study is the question of
re-entrance effects-- is the behaviour of the conductance simply
monotonic in $V$ and $B$?  In the case of a diffusive NS junction,
it is well known that this is not the case \cite{NAZAROV,CHARLAT}.
At zero voltage and also at voltages large enough to break
electron-hole degeneracy, the conductance of the junction is the
same as in the normal state. However, a conductance enhancement
does occur at intermediate voltages. We examine re-entrance
effects in the conductance of chaotic Andreev billiards, both in the case of
ballistic contacts, and in the case where both point contacts
contain opaque tunnel barriers.

The remainder of this paper is organized as follows.  In Section
II, we present our model for the chaotic dot
and introduce the technique used to calculate the voltage and field-dependent
Andreev conductance. In
Section III, we discuss results in the case where the contact to
the superconductor dominates the contact to the normal electrode,
and contrast the results for a chaotic billiard with exact quantum
mechanical calculations for two integrable systems. In Section IV,
we discuss re-entrance effects.
We conclude with a synopsis of our key results
in Section V.

\section{Model and Technical Details}

\subsection{Formulation of the Problem}

\begin{figure}
\centerline{\psfig{figure=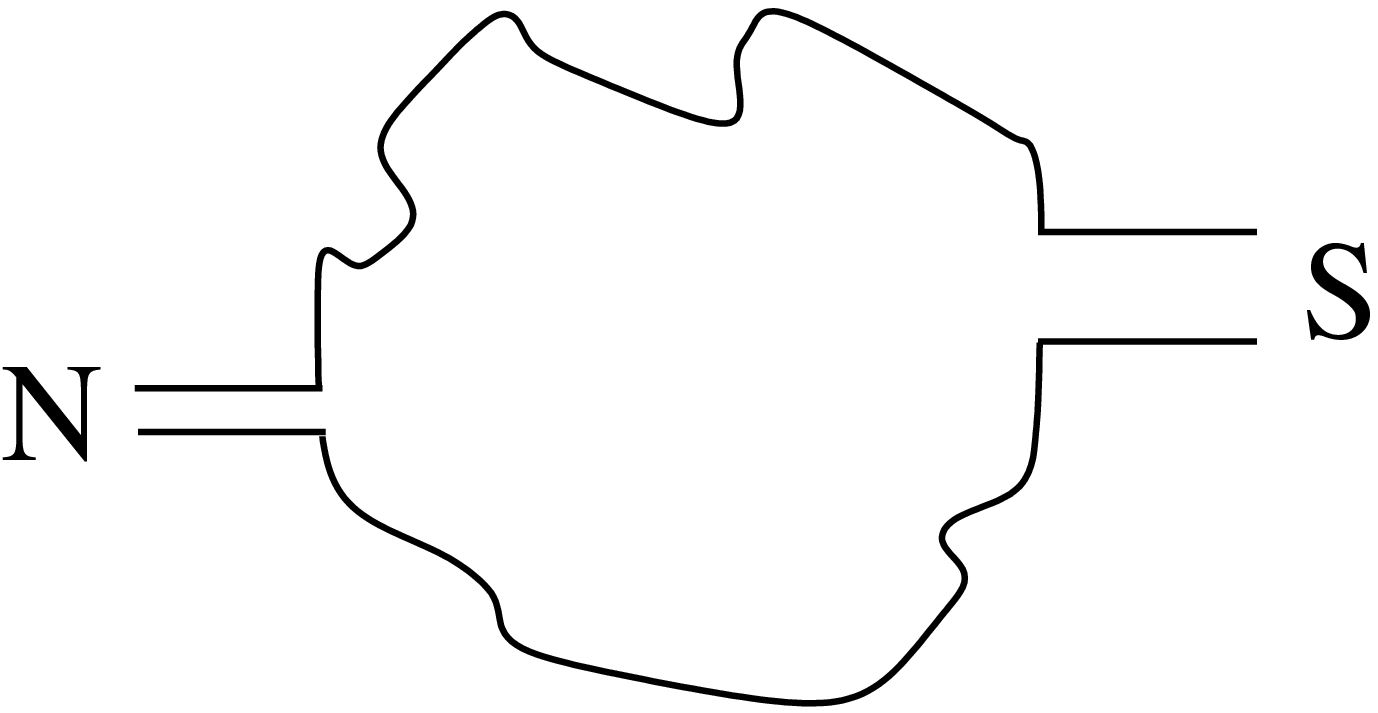,width=4.8cm}}
\refstepcounter{figure}
\label{CDotFig}
\bigskip
{
\small FIG. \ref{CDotFig}.
Schematic drawing of a dot coupled via point contacts to a normal
metal and a superconductor.
}
\end{figure}

We consider a chaotic quantum dot coupled
via point contacts to a normal
metal and a superconductor, each having respectively $N_N$ and
$N_S$ propagating modes at the Fermi energy $E_F$ (see Fig. \ref{CDotFig}).
We assume
that the ergodic time is much shorter than
other relevant timescales of the dot (i.e. the inverse superconducting
gap $\hbar / \Delta$ and the dwell time), so that random
matrix theory (RMT) \cite{Mehta} may be used to describe its transport and
spectral properties.   In RMT, the Hamiltonian of dot is represented
by a $M \times M$ Hermitian matrix $H$ which, at zero magnetic
field, is real symmetric and a member of the Gaussian orthogonal
ensemble:
\begin{equation}\label{GOEDefn}
    P(H) = \exp(-\frac{1}{4} M \lambda^{-2} \text{tr} H^2).
\end{equation}
The matrix size $M$ is sent to infinity at the end of the
calculation.  The energy scale $\lambda$ is related to the mean
level spacing $2 \delta$ by $\lambda = \frac{2 M \delta}{\pi}$.
More specifically, $2 \delta$ is the mean level spacing for
particle-like excitations in the absence of a coupling to the
superconductor; with the superconductor, the relevant excitations
are of mixed particle-hole type, and have a level spacing
$\delta$ for excitation energies much larger than the gap energy.

In the case of a non-zero magnetic field, the Hermitian matrix $H$
is a member of the Pandey-Mehta distribution \cite{Mehta,Pandey}:
\begin{equation}
\label{PandeyDist}
 P(H) \propto  e^{
-\frac{M(1+\gamma^2)}{4\lambda^2} \Sigma_{i,j=1}^{M} \left[
(\text{Re}H_{ij})^2 + \gamma^{-2}(\text{Im}H_{ij})^2 \right]}
\end{equation}
As $\gamma$ increases from $0$ to $1$, the distribution evolves
from one with complete time-reversal symmetry (Gaussian orthogonal
ensemble) to one with no time-reversal symmetry (Gaussian unitary
ensemble). The parameter $\gamma$ may be related to the magnetic
flux $\Phi$ through a two-dimensional dot having area $A$
\cite{FluxRefA,FluxRefB}:
\begin{equation}\label{FluxReln}
  M \gamma^2 = C \left(\frac{\Phi}{\Phi_0}\right)^2 \frac{\hbar
  v_F}{\sqrt{A} \delta},
\end{equation}
where $\Phi_0 = hc/e$ is the flux quantum, $v_F$ is the Fermi
velocity, and $C$ is a constant of order unity.

The normal $(N_N + N_S) \times (N_N+N_S)$ scattering matrix
$S(\ve)$ of the system at an energy $\ve$ above $E_F$ can be
expressed in terms of the matrix $H$ \cite{VWZ}:
\begin{eqnarray}\label{SDefn}
    S(\ve) & = & 1 - 2 \pi i W^{\dag}\left(\ve - H - i \pi W
    W^{\dag}\right)^{-1} W  \\
  \nonumber
  & = & \left( \begin{array}{cc}
    r_{NN}(\ve) & t_{NS}(\ve) \\
    t_{SN}(\ve) & r_{SS}(\ve) \
  \end{array}   \right),
\end{eqnarray}
where $W$ is an $M \times (N_N + N_S)$ matrix representing the
coupling between the point contacts and the dot, having elements:
\begin{eqnarray}\label{WDefn}
    W_{mn}  & = & \delta_{mn}\frac{1}{\pi} \left(2M\delta\right)^{1/2}
                \left(
                    \frac{2 - T_n -2 \sqrt{1-T_n} }{T_n}  \right)^{1/2}  \\
     & = &
    \delta_{mn}\left( \frac{\lambda}{\pi} Z_n \right)^{1/2}
    \nonumber
\end{eqnarray}
The $T_n$ are the transmission probabilities for each mode, which
we take simply as $T_N$ for modes coupled to the normal electrode,
and $T_S$ for those coupled to the superconducting electrode (this
in turn defines $Z_N$ and $Z_S$).

For voltages below the excitation gap $\Delta$ of the
superconductor, electrons and holes incident on the
dot-superconductor interface may be Andreev reflected.  In this
case scattering from the dot, as seen from the normal metal
contact, can be represented by a $2N_N \times 2N_N$ scattering
matrix $\mathcal{S}$:
\begin{equation}\label{SuperSDefn}
  \mathcal{S} =  \left(
                    \begin{array}{cc}
                      r_{ee}(\ve) & r_{eh}(\ve) \\
                      r_{he}(\ve) & r_{hh}(\ve)
                    \end{array}
                    \right).
\end{equation}
Here, $r_{he}(\ve)$ is the $N_N \times N_N$ matrix describing the
Andreev reflection of an incoming electron in the N point contact
to an outgoing hole in the same lead; $r_{eh}(\ve)$, $r_{ee}(\ve)$
and $r_{hh}(\ve)$ are defined analogously.  These matrices may be
written in terms of the sub-matrices of the normal scattering
matrices $S(\ve)$ \cite{BeenRMT}.  For $r_{he}(\ve)$ one has:
\begin{equation}\label{RheDefn}
  r_{he}(\ve) = t_{NS}^{*}(-\ve) M_{ee}(\ve)
  \alpha(\ve)t_{SN}\vphantom{*}(\ve),
\end{equation}
with
\begin{eqnarray}\label{MeeDefn}
    \nonumber
  M_{ee}(\ve) & = &  \left[(1-\alpha(\ve)^2 r_{SS}^{\phantom{*}}(\ve)
  r_{SS}^{*}(-\ve)\right]^{-1} \\
    \nonumber
 \alpha(\ve) & = & \exp(-i \arccos(\ve/\Delta))
\end{eqnarray}

For voltages below the gap $\Delta$ in the S electrode, the zero
temperature conductance of the system is given by the
Tabikane-Ebisawa formula \cite{Tabikane}:
\begin{equation}\label{TEFormula}
  G(eV) = \frac{4e^2}{h} \tr r_{he}^{\dag}(eV)
    r_{he}^{\phantom{\dagger}}(eV).
\end{equation}

We wish to calculate the ensemble-averaged Andreev conductance of
the dot for arbitrary values of voltage $V$ and magnetic field
$B$; previous studies focused exclusively on the cases where $V,B$
were $0$ or large enough to completely break the symmetry between
electrons and holes.  To this end, we first rewrite Eq.
(\ref{SuperSDefn}) in a manner which is formally equivalent to the
normal-state expression (\ref{SDefn}).  Letting $\Omega_S = \pi
W_S W^{\dag}_S$ and $\Omega_N = \pi W_N W^{\dag}_N$, we define the
$2M \times 2M$ effective particle-hole Hamiltonian
\cite{FRAHM,BROUWERJJ}:
\begin{equation}\label{HHdefn}
  \HH  =  \left(
            \begin{array}{cc}
                H & 0 \\
                0 & -H^{*}
            \end{array}
        \right),
\end{equation}
and the self-energies from the leads:
\begin{eqnarray}\label{SEDefn}
    \nonumber
  \Sigma^{0}_{N}(\ve) & = & -i
        \left(
            \begin{array}{cc}
                \Omega_N     &   0  \\
                0                  &   \Omega_N
            \end{array}
        \right)             \\
  \Sigma^{0}_{S}(\ve) & = &  -  \frac{\Delta}{\sqrt{\Delta^2-\ve^2}}
        \left(
            \begin{array}{cc}
                (\ve/\Delta) \Omega_S     &   \Omega_S  \\
                \Omega_S                  &   (\ve/\Delta) \Omega_S
            \end{array}
        \right)
\end{eqnarray}
With these definitions in place, a direct algebraic manipulation
shows that the particle-hole scattering matrix $\SS$ in
(\ref{SuperSDefn}) can be expressed in terms of the effective
retarded Green function $\GG_R$,
\begin{equation}\label{GGDefn}
  \GG_R(\ve) = \left[ \ve - \HH(\ve) - \Sigma^{0}_N(\ve) - \Sigma^{0}_S(\ve)
  \right]^{-1},
\end{equation}
by
\begin{eqnarray}\label{SSDefn}
    \SS(\ve) & = & 1 - 2 \pi i
                \left(
                    \begin{array}{cc}
                        W_N^{\dag}  &   0   \\
                        0           &   W_N^{\dag}
                    \end{array}
                \right)
                \GG_R(\ve)
                \left(
                    \begin{array}{cc}
                        W_N  &   0   \\
                        0           &   W_N
                    \end{array}
                \right)
\end{eqnarray}

 The major simplification offered by Eq.
(\ref{SSDefn}) compared to Eq. (\ref{SuperSDefn}) is that it
allows one to compute the conductance in a manner analogous to
that used in the normal state, as we now demonstrate.  From this
point onward, we focus on the case $\Delta \gg \ve, eV$.
In this regime, the properties of the
system become independent of the specific details of the
superconductor.  Calculations retaining a finite $\Delta$ will be
presented elsewhere \cite{AASH}.

\subsection{Details of the Calculation - $B = 0$}

\begin{figure}
\centerline{\psfig{figure=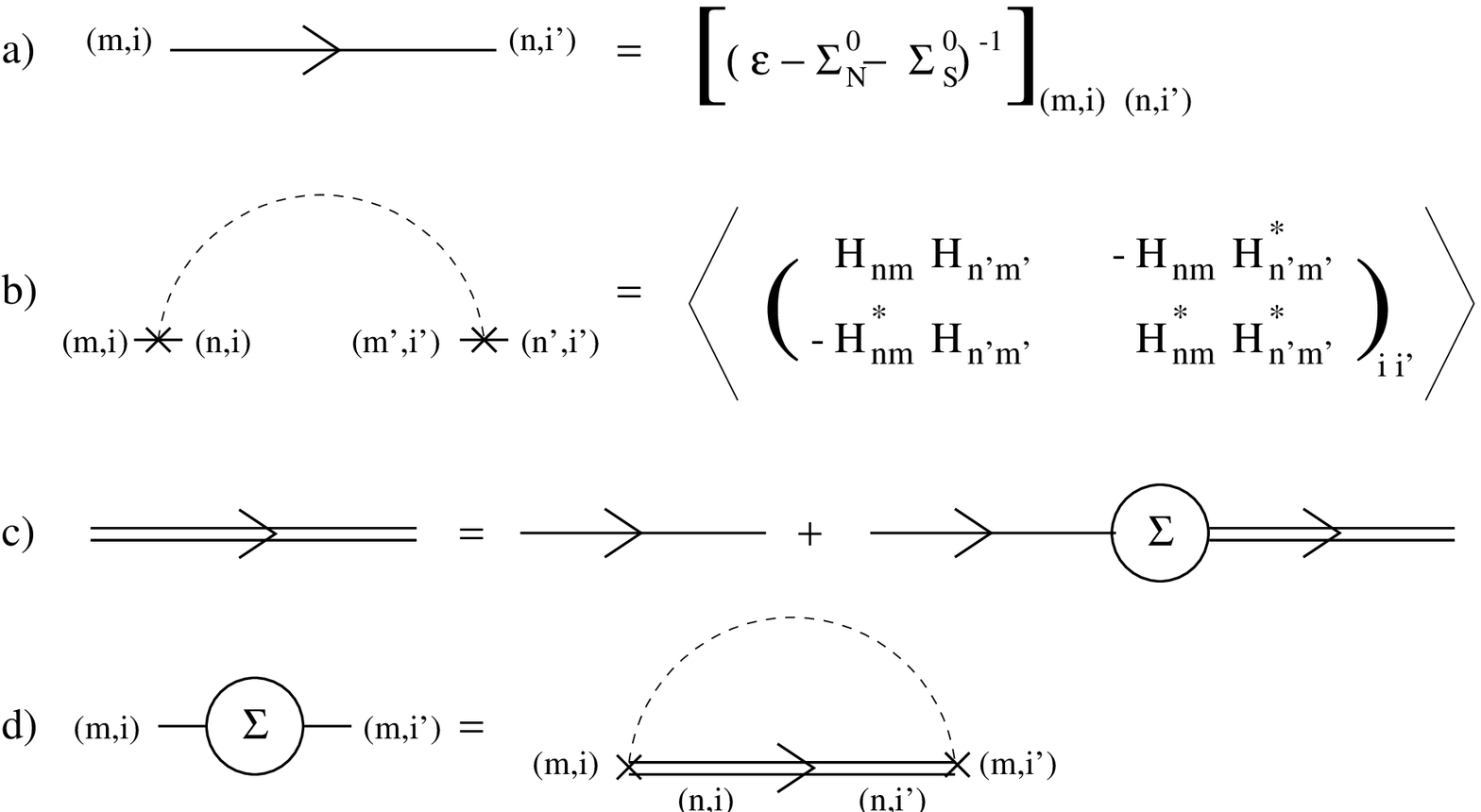,width=8.5cm}}
\refstepcounter{figure}
\label{SelfEnergFig}
\bigskip
{
\small FIG. \ref{SelfEnergFig}.
(a) Definition and diagrammatic representation of the unaveraged
Green function.  This is a $2M\times 2M$ matrix; $n,m$ refer to mode
indices, while $i,i'$ are electron-hole indices which can take one
of two values ($e$ or $h$).
(b) The effective Hamiltonian $\HH$ is represented by a
cross; the dashed lines indicates averaging.  (c) Diagrammatic
Dyson equation; the double line is the averaged Green function
$\langle \GG \rangle$.
(d) Self energy to leading order in
$1/M$; the neglected higher order terms correspond to
graphs with crossed dashed lines.  The intermediate index $m$
is to be summed over.  Note that $\Sigma$ and hence $\langle \GG \rangle$
are diagonal in mode space but have off-diagonal terms
in particle-hole space.
}
\end{figure}

\begin{figure}
\centerline{\psfig{figure=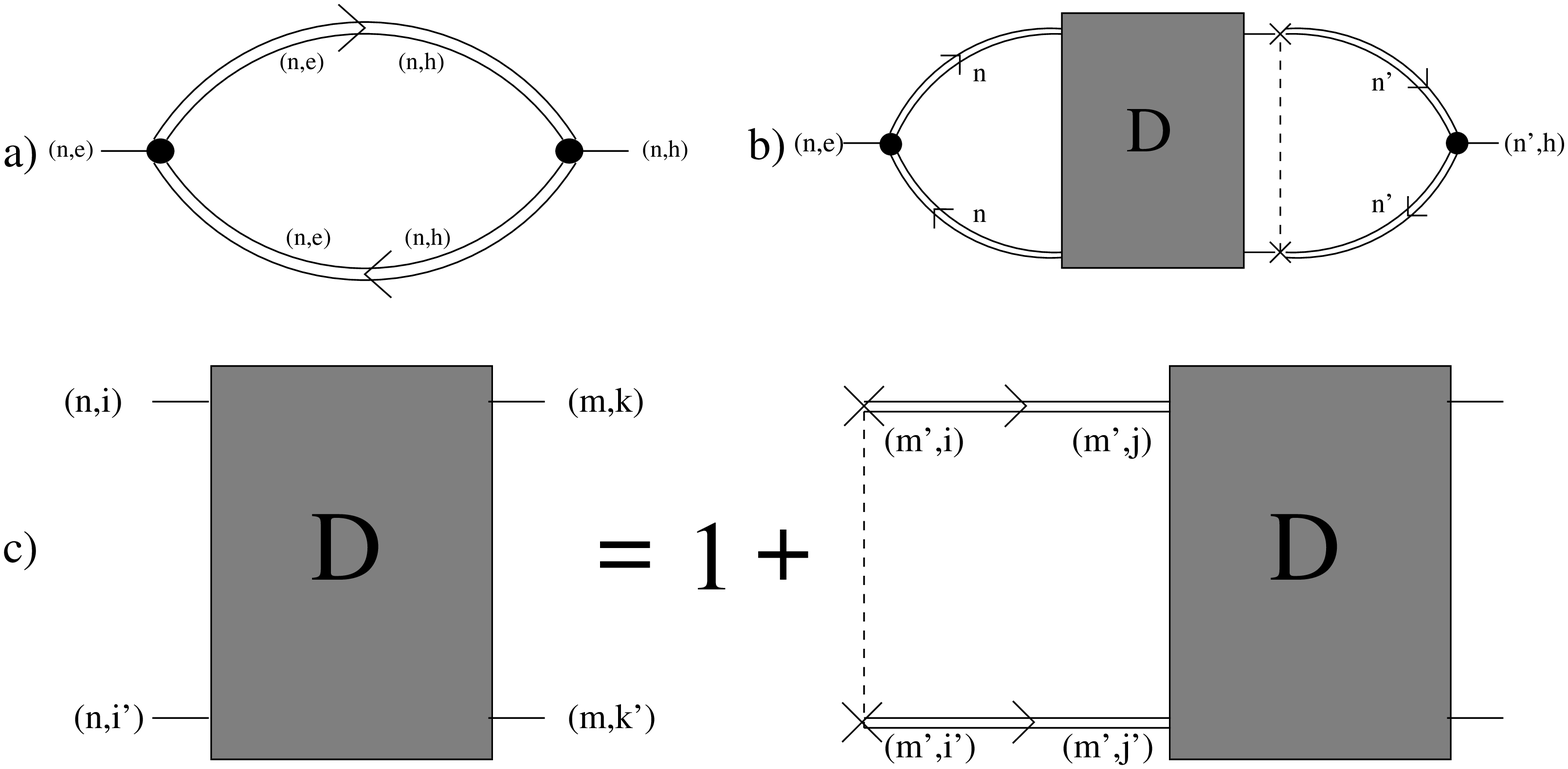,width=8.5cm}}
\refstepcounter{figure}
\label{DiffusonFig}
\bigskip
{\small FIG. \ref{DiffusonFig}. (a) direct contribution to the
Andreev conductance. The trace appearing in Eq. (\ref{TEFormula})
is represented by the bubble diagram.  The double lines indicate
the averaged Green function $\langle \GG \rangle$.  The index $n$
corresponds to a mode coupled to the normal contact and should be
summed over. (b) Diffusion contribution to the Andreev
conductance. (c) Dyson equation for the $4M \times 4M$ matrix D.
Note that upper and lower branches need to have matching mode
indices, but not matching particle-hole indices. }
\end{figure}

We proceed to average Eq. (\ref{TEFormula}) for the Andreev
conductance over the Gaussian orthogonal ensemble defined by Eq.
(\ref{GOEDefn}), which is the appropriate ensemble for zero
magnetic field. We make use of the relation
\begin{equation}\label{avgform}
  \langle H_{ij} H_{kl} \rangle =
    \bigl( \delta_{ik}\delta_{jl} + \delta_{il}\delta_{jk} \bigr)
    \frac{\lambda^2}{M}.
\end{equation}
By expressing the scattering matrix in terms of the Green function
$\GG$, the trace appearing in the conductance formula may be
represented as a standard ``conductance bubble'' diagram (see Fig.
\ref{DiffusonFig}). Further, at this stage of the calculation $\lambda^2/M$ is
taken to be a small parameter, meaning that the usual diagrammatic
technique for impurity averaging may be used.   To obtain the
leading-order result in $1/N_N$,$1/N_S$, we need only sum diagrams which
have no crossed lines (lines indicate the averaging of matrix
elements of $H$ using the rule (\ref{avgform}); see Fig. \ref{SelfEnergFig}).
Note that $\lambda$ will be sent to infinity only at the very end
of the calculation; this procedure corresponds to first taking the
matrix size to infinity while keeping a finite bandwidth, and then
{\it independently} taking the bandwidth to infinity. This
ordering of limits is necessary to generate a well-defined
perturbative expansion.

The first step in this framework is to calculate the averaged
matrix Green function $\langle \GG \rangle$.  As in
\cite{MELSEN,BROUWERJJ}, we find the following
self-consistent Dyson equation:
\begin{mathletters}
\begin{equation}\label{GDyson}
  \langle \GG(\ve) \rangle = \left[ \ve - \Sigma^{0}_N(\ve) -
                                \Sigma^{0}_S(\ve) - \Sigma(\ve)
                            \right]^{-1},
\end{equation}
where the self-energy from averaging $\Sigma$ is given by
\begin{eqnarray}
    \nonumber
    \Sigma & = & 1_M \otimes \left( \begin{array}{cc}
                            \Sigma_{ee} & \Sigma_{eh} \\
                            \Sigma_{he} & \Sigma_{hh}
                        \end{array} \right)          \\
    \label{AvgSE}
               & = &  1_M \otimes \frac{\lambda^2}{M}
                            \left( \begin{array}{cc}
                            \langle \tr \GG_{ee}\rangle & -\langle \tr \GG_{eh}\rangle \\
                          -\langle \tr \GG_{he}\rangle & \langle \tr \GG_{hh}\rangle
                        \end{array} \right).
\end{eqnarray}
\end{mathletters}%
In this equation, $1_M$ denotes the $M \times M$ unit matrix, and
$\otimes$ denotes a direct product; it is pictured diagrammatically
in Fig. (\ref{SelfEnergFig}).  Having performed the summation of
diagrams, we now let $M \rightharpoonup \infty$ keeping $\ve$ ,
$\delta$ , $N_S$ and $N_N$ fixed.

Three of the equations thus obtained relate the components of the
self energy to one another:
\begin{mathletters}
\begin{equation}\label{ASE1}
    \See = \Shh \text{      ,      } \Seh = \She,
\end{equation}
\begin{equation}\label{ASE2}
  \Seh^2 - \See^2 = \lambda^2.
\end{equation}
\end{mathletters}%
The last equation allows us to parameterize $\Sigma$ in terms of a
pairing angle $\theta(\ve)$:
\begin{mathletters}
\begin{eqnarray}\label{trig}
    \lambda \sin\theta(\ve) & = & -\Seh(\ve),  \\
    \lambda \cos\theta(\ve) & = & i \See(\ve).
\end{eqnarray}
\end{mathletters}%
The remaining self-energy equations now take the form:
\begin{mathletters}
\label{SelfEnEqs}
\begin{eqnarray}\label{ASE3}
    \tan(\theta(\ve)) & = & \frac{N_S Q_S}{N_N Q_N -i \frac{\pi \ve}{2
    \delta} },  \\
\label{ASE4}
    Q_N  & = &  \frac{T_N}{2} \big( 2 - T_N \left[
    1 - \cos \theta(\ve)  \right]
                                \big)^{-1},  \\
\label{ASE5}
    Q_S  & = &   \frac{T_S}{2} \big( 2 - T_S \left[
    1 + \sin \theta(\ve)  \right]
                                \big)^{-1}.
\end{eqnarray}
\end{mathletters}%
To obtain a unique solution of the self-energy equations we have
imposed the boundary condition $\See(\ve) \rightharpoonup -i
\lambda$ as $\ve \rightharpoonup \infty$; this represents the
physical condition that we recover a normal metal with a constant
density of states at large energies.  The pairing angle $\theta$
can is related the density of states of the dot:
\begin{equation}\label{DOSdefn}
  \rho(\ve) = -\frac{\text{Im}}{\pi} \text{tr } \langle \GG(\ve+0^{+})
  \rangle = \frac{\text{Re} [ \cos(\theta(\ve)) ] }{\delta};
\end{equation}
$\theta = \pi/2$ corresponds to a fully superconducting state,
while $\theta = 0$ corresponds to the normal state. It is
interesting to note the analogy between $\theta(\ve)$ and the
pairing angle used in the quasi-classical theory of dirty
normal-superconducting interfaces \cite{QUASICLASSICAL} and in the
circuit theory of Andreev conductance \cite{CIRCUITS} . In the
present case, the normalization condition (\ref{ASE2}) does not
need to be externally imposed, but is a direct consequence of the
averaging procedure.

While the leading order solution of the self energy $\Sigma$ is
sufficient if one is only interested in the density of states (as
in ref. \cite{BROUWERJJ}), the conductance calculation requires
that $N_S/M$ , $N_N/M$ and $\ve / M \delta$
corrections to Eq. (\ref{ASE2}) be calculated.  This
correction may be expressed in terms of the leading order self
energy solution:
\begin{eqnarray}\label{SENorm}
\frac{\Seh^2 - \See^2}{\lambda^2} - 1  & = &
   -\frac{1}{M} \bigg(N_S Q_S \left(\sin(\theta) + Z_S \right) \\
& & \mbox{} + N_N Q_N \left(\cos(\theta) + Z_N \right)
          + i \frac{\pi \ve}{2 \delta}
                    \bigg)
                    \nonumber
\end{eqnarray}

Having computed the self-energy from averaging $\Sigma$ and thus
the average Green function $\langle \GG \rangle$, we can proceed
to sum diagrams for the conductance. In analogy to the usual
impurity technique \cite{MAHAN}, two sets of contributions arise
(see Fig. \ref{DiffusonFig}a) to leading order: $G = (4e^2/h)(
g_{\rm Dir} + g_{\rm Diff})$. The first is a direct contribution
which is completely determined by $\langle \GG \rangle$, that is,
by the ensemble-averaged probability amplitude $\langle r_{he}
\rangle$:
\begin{eqnarray}
g_{\rm Dir} & = &  (2 \pi)^2 \text{tr }
  W_N^{\dagger} \langle \GG \rangle W_N^{\phantom{\dagger}}
  W_N^{\dagger} \langle \GG \rangle^\dagger W_N^{\phantom{\dagger}}
  \nonumber \\
           & = &    \text{tr } \langle r_{he} \rangle
               \langle r_{he} \rangle^\dagger.
\label{HeurGDrude}
\end{eqnarray}
This contribution may be interpreted as arising from Andreev
reflections which effectively occur at the interface of the normal
lead and the cavity. The second contribution is due to the
fluctuations of $r_{he}$:
\begin{equation}
\label{HeurGDiff} g_{\rm Diff} =   \left\langle
    \left(r_{eh} - \langle r_{eh} \rangle \right)
    \left(r_{eh} - \langle r_{eh} \rangle \right)^2  \right \rangle.
\end{equation}
It describes the current carried by quasiparticles in the dot
which are Andreev reflected at the interface of the dot and the
superconductor.  Diagrammatically, it is equivalent to a diffusion
ladder, where the averaging links the upper and lower branches of
the conductance bubble (see Fig. \ref{DiffusonFig}b-d)
\cite{DECOMPNOTE}.  We ignore here quantum corrections, which are
formally smaller by a factor of $\max(1/N_N,1/N_S)$. Note that in
computing the diffusion sum, each individual graph is of order
$1/M$; this means that terms of order $1/M$ must be retained when
computing the $4 \times 4$ matrix inverse arising from summing the
series.  As with the calculation of the average Green function
$\langle \GG \rangle$, we only let $M$ tend to infinity after
having performed the partial summation of diagrams.

To present the results of the conductance calculation, we first
define the following kernel functions:
\bleq
\begin{mathletters}
\begin{eqnarray}\label{LambdaKernel}
    \Lambda(\ve) & = & 2 N_N
        \left(
            1 - \left| \frac{Q_N}{\GN} \right|^2
            \left| 1 + \GN \cos(\theta) \right|^2
        \right)  +   2 N_S  \left(
            1 - \left| \frac{Q_S}{\GS} \right|^2
            \left| 1 + \GS \sin(\theta) \right|^2
        \right)  +                \\
    \nonumber
    & & \mbox{} -2 \text{Re} \Bigl[ N_N Q_N \left(\GN + \cos(\theta)\right) +
        N_S Q_S \left(\GS + \sin(\theta)\right) \Bigr] +
        \frac{\pi \ve}{\delta} \text{Im } [\cos(\theta)] ,    \\
    \label{OmegaKernel}
    \Omega(\ve) & = & 2 N_N \bigl| Q_N \sin(\theta) \bigr|^2 +
                      2 N_S \bigl | Q_S \cos(\theta) \bigr|^2,   \\
    \label{PiNKernel}
    \Pi_N(\ve) & = & 2 N_N |Q_N |^2 \Bigl\{
        \left( 1 + 6\GN^2 + \GN^4 \right) \left(1 + |\cos(\theta)|^2 \right)
        + 8 \GN \left(1+\GN^2\right) \text{Re} \left[ \cos(\theta) \right]
    \Bigr\}, \\
    \label{PiSKernel}
    \Pi_S(\ve) & = &  2 N_S |Q_S |^2 \Bigl\{
        \left( 1 + 6\GS^2 + \GS^4 \right)
         \left( 1 + |\cos(\theta)|^2 \right)
        + 8 \GS \left(1+\GS^2\right) \text{Re} \left[ \cos(\theta) \right]
    \Bigr\},
\end{eqnarray}
\end{mathletters}%
\eleq \noindent where $Z_N$ and $Z_S$ were defined below Eq.
(\ref{WDefn}). With these definitions, we find:
\begin{mathletters}
\begin{equation}\label{gDrude}
  g_{\rm Dir} =  2 N_N
  \frac{T_N^2}{\left| 2 - 2 T_N \sin^2(\theta/2) \right|^2}
  \left| \sin(\theta) \right|^2,
\end{equation}
\begin{eqnarray}
\label{gD1}
g_{\rm Diff} & = & 8 N_N^2 \left| \frac{Q_N^2}{Z_N} \right|^2
  \left(\Lambda^2 - \Omega^2 \right) ^{-2} \bigg(\Pi_N |\sin(\theta)|^2 \\
& &  \mbox{} + \left(1 - Z_N^2 \right)^2 \Lambda | \sin(\theta)
|^2 + \Pi_S |\cos(\theta)|^2 \bigg).
\nonumber
\end{eqnarray}
\end{mathletters}%
Equations (\ref{LambdaKernel}) - (\ref{gD1}), along with
Eq.(\ref{SelfEnEqs}) for the self energy $\Sigma$, determine the
ensemble-averaged $B=0$ Andreev conductance through the dot to
leading order in $1/N_N$ , $1/N_S$ for all voltages such that $eV
\ll \Delta$.  The extension of these formulas to non-zero $B$ are
presented in the Appendix.

\section{Probing Induced Superconductivity via Conductance}

\subsection{Chaotic Dot}

In this section, we consider the situation where the coupling
between the dot and the superconductor is much stronger than the
coupling between the dot and the normal lead, $N_N T_N \ll N_S
T_S$. In this limit,  the normal metal will only weakly perturb
the properties of the dot-superconductor system; thus, we may view
the conductance through the structure as being a probe of the
induced superconductivity in the dot. As mentioned in the
Introduction, previous studies have examined the density of states
of such Andreev billiards in the absence of a normal lead
\cite{MELSEN}. It was found for the case of a chaotic dot that an
energy gap $\IGap$ opens up on a scale set by the inverse of the
time needed for a particle to escape from the dot to the
superconductor:
\begin{eqnarray}
\label{IndDelta}
  \IGap & = &   c (E_S)    \\
\label{ESDefn}
 E_S & = & \left(\frac{\hbar}{\tau_{\rm S. esc}}\right)
                 = \left(\frac{N_S T_S \delta}{2\pi}\right)
\end{eqnarray}
The parameter c is of order unity and a monotonic function of
$T_S$; it varies from $0.6$ in the case of no tunnel barrier, to
$1$ in the case of an opaque tunnel barrier. It was also found
that the shape of the density of states above $\IGap$ was vastly
different in these two limits:  in the tunnel regime $T_S\ll1$, the
density of states was BCS-like, having a square-root singularity
at $\IGap$, whereas in the case of no tunnel barrier, the density
of states gradually increased above $\IGap$.  Note that the gap
$\IGap$ does not depend on the gap $\Delta$ in the bulk
superconductor, which was taken to infinity in the calculations.

The question that naturally arises here is how the induced
superconductivity seen in the density of states manifests itself
in the conductance.  Some insight may be obtained by considering
equations (\ref{gDrude}) and (\ref{gD1}) to lowest order in the
small parameter $\eta = N_N T_N / N_S T_S$.  Without the normal
lead, we have a gap in the dot density of states up to an energy
$\IGap$. As the density of states is proportional to $\text{Re}
\left[\cos(\theta(\ve))\right]$, this implies that $\text{Re}
\left[\theta(\ve)\right]=\pi/2 + O(\eta)$ for $\ve < \IGap$ .
Using this, the ``sub-gap'' (i.e. $\ve < \IGap$) direct contribution
to the conductance takes the form:
\begin{eqnarray}\nonumber
  g = g_{\rm Dir}(\ve) & = &
    \frac{2 N_N T_N^2}{(2-T_N)^2 - 4(1-T_N)(\text{Im}
    [\sin(\theta_0(\ve))])^2}  \\
\label{BTK} & & \mbox{}+ O(\eta),
\end{eqnarray}
while the diffusion contribution $g_{\rm Diff}$ is of order
$\eta^2$ and thus negligible.  As the energy is increased above
the gap, the density of states returns to its normal-state value,
and consequently $\theta(\ve) \rightharpoonup 0$ to leading order.
For $\ve \gg \IGap$ the direct term is negligible (being
proportional to $\sin(\theta)$), while the diffusion term gives
the normal-state conductance:
\begin{equation}\label{normcond}
   g = g_{\rm Diff}(\ve) = N_N T_N + O(\eta).
\end{equation}

The above considerations become extremely suggestive when one
considers the case $T_N = 1$.  There is a perfect conductance
doubling for voltages below the effective gap $\IGap$, while at
higher voltages the conductance drops to its normal state value:
\begin{equation}\label{gperfect}
    g = \cases{
        2 N_N,  &for $\ve \leq \IGap$\cr
        N_N, &for $\ve \gg \IGap$.\cr}
\end{equation}

\begin{figure}[t]
\centerline{\psfig{figure=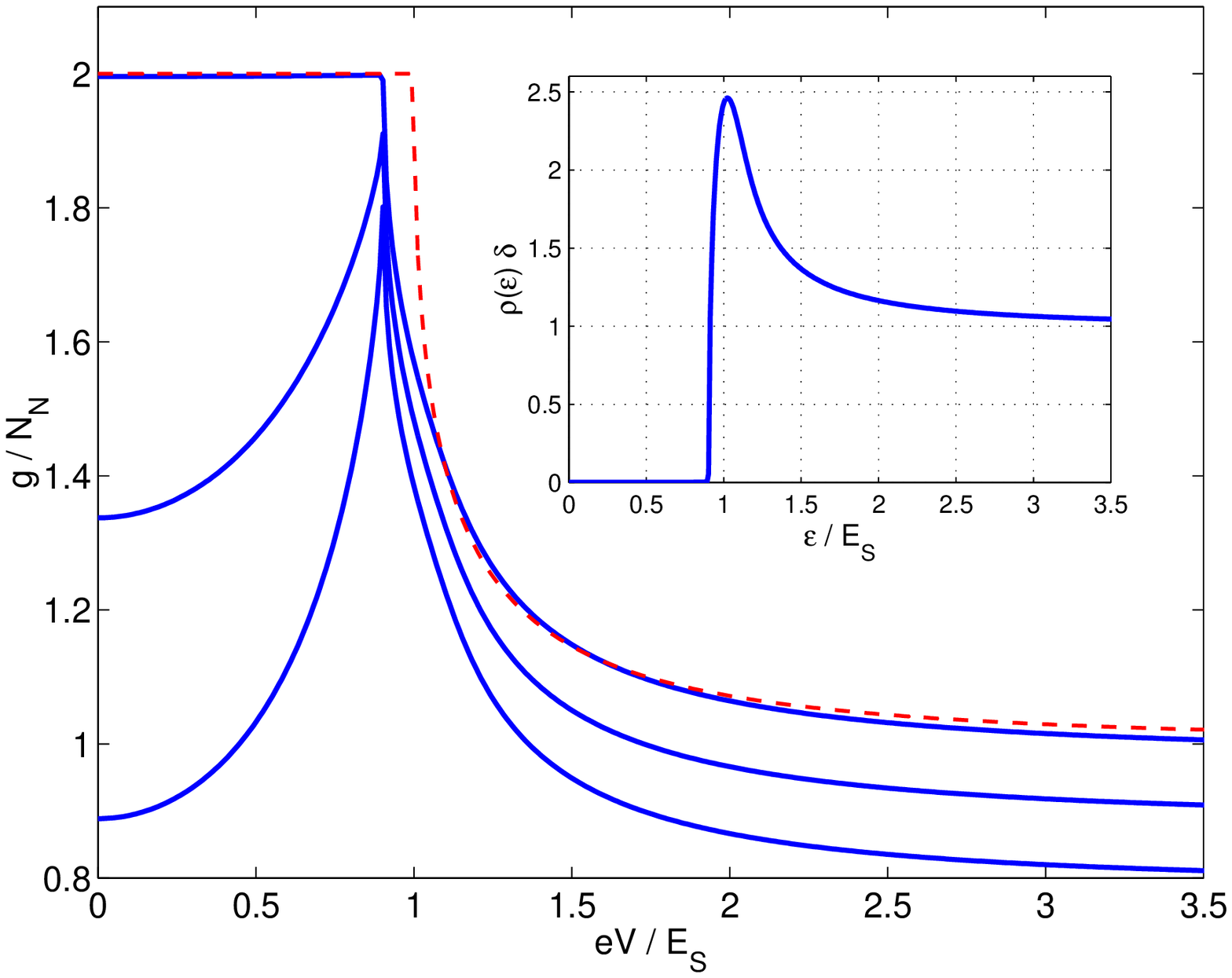,width=8.5cm}}
\refstepcounter{figure} \label{TIndfig} { \small FIG.
\ref{TIndfig}. Conductance versus voltage for $N_S = 10^4 N_N$,
$T_N = 0.1$, illustrating the ``local Andreev reflection'' effect,
in which the dot itself acts as a superconductor having a gap
$\IGap$.  Solid lines are calculated curves for, from top to
bottom $T_N = 1,0.9,0.75$. The dashed curve shows the BTK result
for $T_N = 1.0$; close agreement is found for all values of $T_N$.
The dot density of states (shown in inset) in this case resembles
a  BCS density of states. The voltage is measured in units of
$E_S$, the inverse of the escape time to the superconductor (cf.
Eq. (\ref{ESDefn})).
 }
\end{figure}

\begin{figure}[t]
\centerline{\psfig{figure=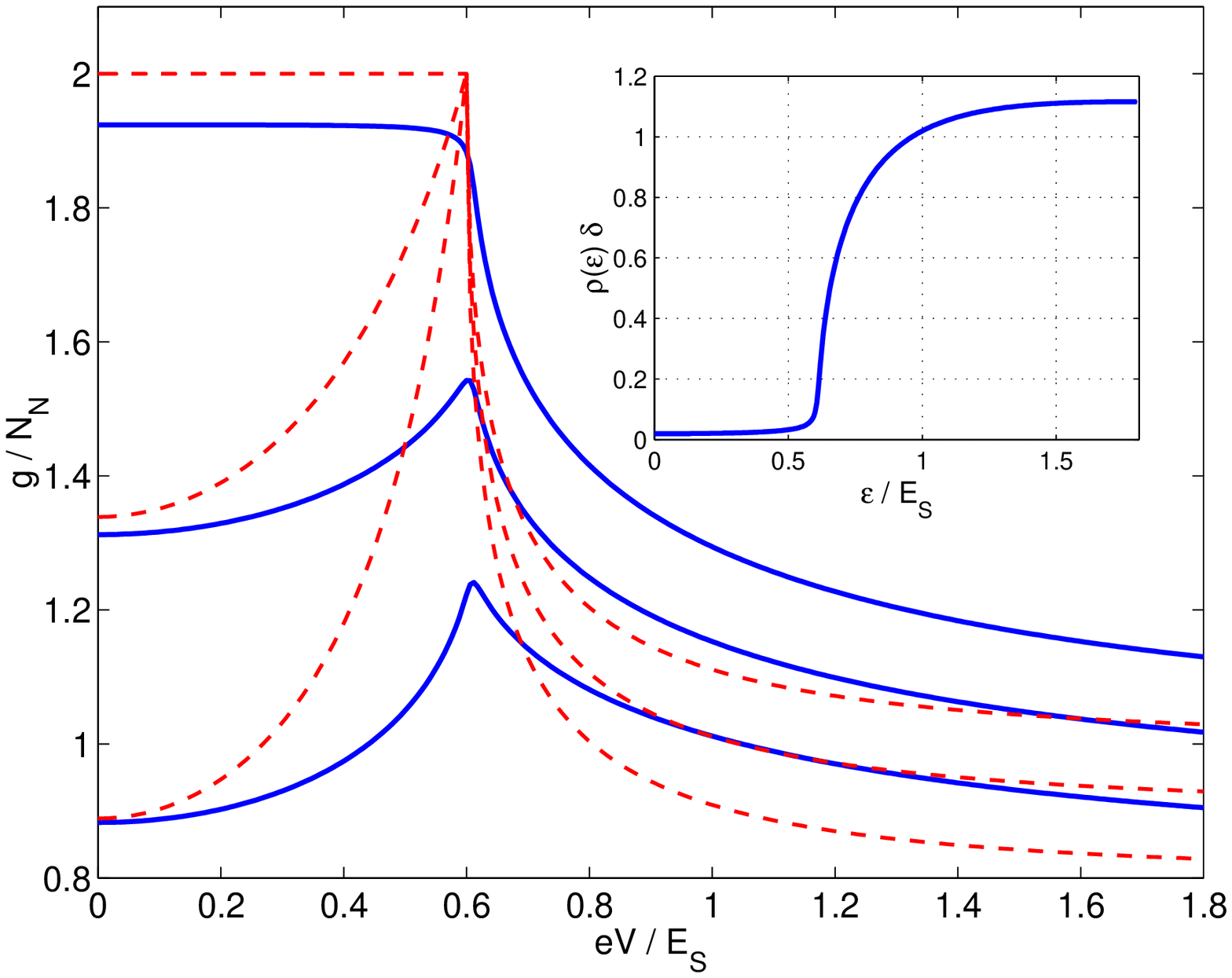,width=8.5cm}}
\refstepcounter{figure} \label{BIndfig} {\small FIG.
\ref{BIndfig}. Conductance versus voltage for $N_S = 10^4 N_N$,
$T_S = 1$, illustrating the ``local Andreev reflection'' effect.
Solid lines are calculated curves for, from top to bottom $T_N =
1,0.9,0.75$; dashed curves show BTK result for same values of
$T_N$. The differences in the dot density of states (shown in
inset) from a BCS density of states leads to large differences
from the BTK lineshapes. }
\end{figure}

This is precisely what would be expected {\it if} the normal point
contact were in perfect contact with a bulk superconductor having
a gap $\IGap$--  conductance doubling would be expected below the
gap due to Andreev reflection, whereas above the gap the
normal-state conductance would be recovered as quasiparticles
would now carry the current.  In this respect, note that below the
induced gap $\IGap$ both the averaged reflection probability
$\langle |r_{eh}|^2 \rangle$ {\it and} the averaged reflection
amplitude  $\left| \langle r_{eh} \rangle \right|$ are non-zero;
for $eV \gg \IGap$, the latter vanishes.  Nonetheless, in both
cases (i.e. for voltages above and below $\IGap$) Andreev
reflection is the only current carrying process.  For energies below
$\IGap$, the dot itself appears as if it were superconducting and Andreev
reflection effectively occurs at the normal lead--dot interface.
For energies far above $\IGap$, current through the dot is
effectively carried by quasiparticles in the dot (which are either
electron or hole-like) which Andreev reflect at the
dot-superconductor interface.

The quantum dot continues to act as if it were a superconductor in
the case where the contact to the normal lead is no longer perfect
(i.e. $T_N \neq 1$); one need only note that if
$\text{Im}[\sin(\theta_0(\ve))]$ is replaced by $\ve/\IGap$,
Equation (\ref{BTK}) becomes identical to the
Blonder-Tinkham-Klapwijk (BTK) formula for the sub-gap conductance
of a normal point contact - superconductor junction, where there
is a tunnel barrier at the interface having transmission $T_N$
\cite{BTK}. The BTK formula is derived assuming there is no
spatial separation between the sites of Andreev scattering and
normal scattering; the only energy dependence thus comes from the
Andreev reflection phase $\alpha(\ve)$, not from the lack of
electron hole degeneracy at finite voltages.  The replacement of
$\ve/\IGap$ by $\text{Im}[\sin(\theta_0(\ve))]$ in Eq. (\ref{BTK})
means that the Andreev reflection phase $\widetilde{\alpha}$ in
the present case is not necessarily the usual $\alpha(\ve)$ given
in Eq. (\ref{RheDefn}), but is rather defined through:
\begin{equation}\label{Ealpha}
\text{Im}[\sin(\theta_0(\ve))]^2 = \frac{1 +
\text{Re}[\widetilde{\alpha}(\ve)^2]}{2}.
\end{equation}
This notwithstanding, the overall implication is still that for
$eV \leq \IGap$, we can effectively consider all Andreev
reflections as occurring at the N-dot interface, as opposed to at
the dot-S interface-- the dot is indeed acting as though it were
itself a superconductor.

We have numerically solved Equations (\ref{ASE3})-(\ref{ASE5}) for
the pairing angle $\theta(\ve)$, and used this to compute the
Andreev conductance versus voltage. In Figures \ref{TIndfig} and
\ref{BIndfig}, we plot the calculated conductance vs. voltage
curves for various values of $T_N$, and compare to what would be
expected from the BTK theory for a simple N-S interface.  We find
an excellent agreement in the case of $T_S \ll 1$ (see Fig.
\ref{TIndfig}) even for voltages above the gap $\IGap$ in the dot.
Such an agreement is not unlikely, as for $T_S \ll 1$ the dot
density of states is BCS-like, and consequently the effective
Andreev phase $\widetilde{\alpha}$ is just equal to the usual
Andreev phase $\alpha$.  In the opposite case of a transparent
contact to the superconductor (Fig. \ref{BIndfig}), clear
deviations from the BTK lineshapes are seen. These deviations
result completely from the fact $\widetilde{\alpha} \neq \alpha$,
which is to be expected as the dot density of states in this case
is quite different from the BCS form.

The ``induced superconductivity'' effect in the dot also manifests
itself in other manners.  A straightforward calculation in the
limit $T_N \rightarrow 0$ shows that the Andreev conductance
becomes proportional to the dot density of states
(i.e.$\text{Re}[\cos(\theta)]$); thus, the Andreev conductance
becomes equivalent to a conventional superconductor
tunneling-density of states measurement.   We have also calculated
the magnetic field dependence of the Andreev conductance in the
limit of small $\eta$ (see Appendix A); here too, our results are
consistent with a picture in which the dot itself acts as a
superconductor. The conductance enhancement at $0$ field remains
constant until a critical flux $\Phi_C$ given by:
\begin{equation}\label{critflux}
  \left( \frac{\Phi_C}{\Phi_0} \right)
    =   C \sqrt{\frac{\IGap}{E_{\rm erg}}}  =  C \sqrt{\frac{\IGap
    \tau_{\rm erg}}{\hbar}}
\end{equation}
where $\tau_{\rm erg}$ is the ergodic time, and C is a
geometry-dependent constant of order unity.  This is the same
critical field required to close the gap $\IGap$.  Note that
unlike a conventional BCS superconductor, where the critical field
is proportional to the gap, in the present case, Eq.
(\ref{critflux}) implies that the critical field is proportional
to $\sqrt{\IGap}$.

The conclusions reached here are markedly different from what
would be expected from a naive trajectory-based semi-classical
analysis \cite{vanWees}. Consider the situation of a ballistic dot
with no tunnel barriers, where $N_S \gg N_N$ . In the semiclassical
picture, electrons entering the dot from the normal lead typically
bounce off the walls of the dot several times before hitting the
superconductor, where they Andreev reflect.  At $V=0$, holes are
the time-reversed partners of electrons; thus when an Andreev
reflection occurs, the hole will retrace the path of the incoming
electron {\it and} cancel its acquired phase. Andreev trajectories
will thus interfere constructively, leading to a large conductance
enhancement at $V=0$ and a non-vanishing average reflection
amplitude $\langle r_{he} \rangle$. This enhancement should be
lost however as the voltage is increased-- at finite $V$,
electrons and holes are no longer degenerate, and the phase
acquired by the hole will not precisely cancel that acquired by
the electron.  The presence of a residual phase $\delta \phi = eVL
/ (\hbar v_F)$ (where $L$ is the length of the trajectory) will
lead to destructive interference, and a consequent decrease in the
conductance. In this picture, even a small voltage should impair
the conductance enhancement seen at zero voltage.

To make this picture quantitative, we may use the fact that our
exact random matrix theory indicates that the conductance below
$\IGap$ is proportional to $| \langle r_{he} \rangle |^2$ (see Eq.
(\ref{BTK})).  Using the semiclassical Andreev phase $\delta \phi
= eVL / (\hbar v_F)$ and the fact that path lengths in chaotic
systems have an exponential distribution function $P(L)=
\exp(-L/\bar{L})/\bar{L} $ \cite{Bauer}, we estimate the average
of the semiclassical reflection amplitude as:
\begin{equation}
\langle r_{he} \rangle_{\rm S.C.} \simeq \int{dL e^{i
\frac{eVL}{\hbar v_F}} P(L) = \frac{1}{1 - i \frac{eV
\bar{L}}{\hbar v_F}} },
\end{equation}
where $\bar{L}$ is the mean path length to the superconductor; for
a ballistic dot, we have $\bar{L} = v_F \tau_{S.esc}$.  Hence:
\begin{equation}
\left| \langle r_{he} \rangle \right|^2 = \left( 1 +
\left(\frac{eV}{c \IGap}\right)^2 \right)^{-1},
\end{equation}
where we have used $\IGap = c \hbar / \tau_{S.esc.}$ (see Eq.
(\ref{IndDelta})) with $c \simeq 0.6$ in the case of a perfect
contact to the superconductor. The semiclassical approach thus
predicts that $\left| \langle r_{he} \rangle \right|^2$ (and hence
the Andreev conductance) will fall off with voltage as a
Lorentzian for  $eV < \IGap$. This is clearly at odds with our
fully quantum mechanic calculation at $T_N = 1$, which shows that
$\left| \langle r_{he} \rangle \right|^2 = 1$ for all voltages
below $\IGap$.  We return to the discrepancy between the
semiclassical and quantum mechanical calculations in the next
subsection.
\subsection{Integrable Dot}

\begin{figure}[t]
\centerline{\psfig{figure=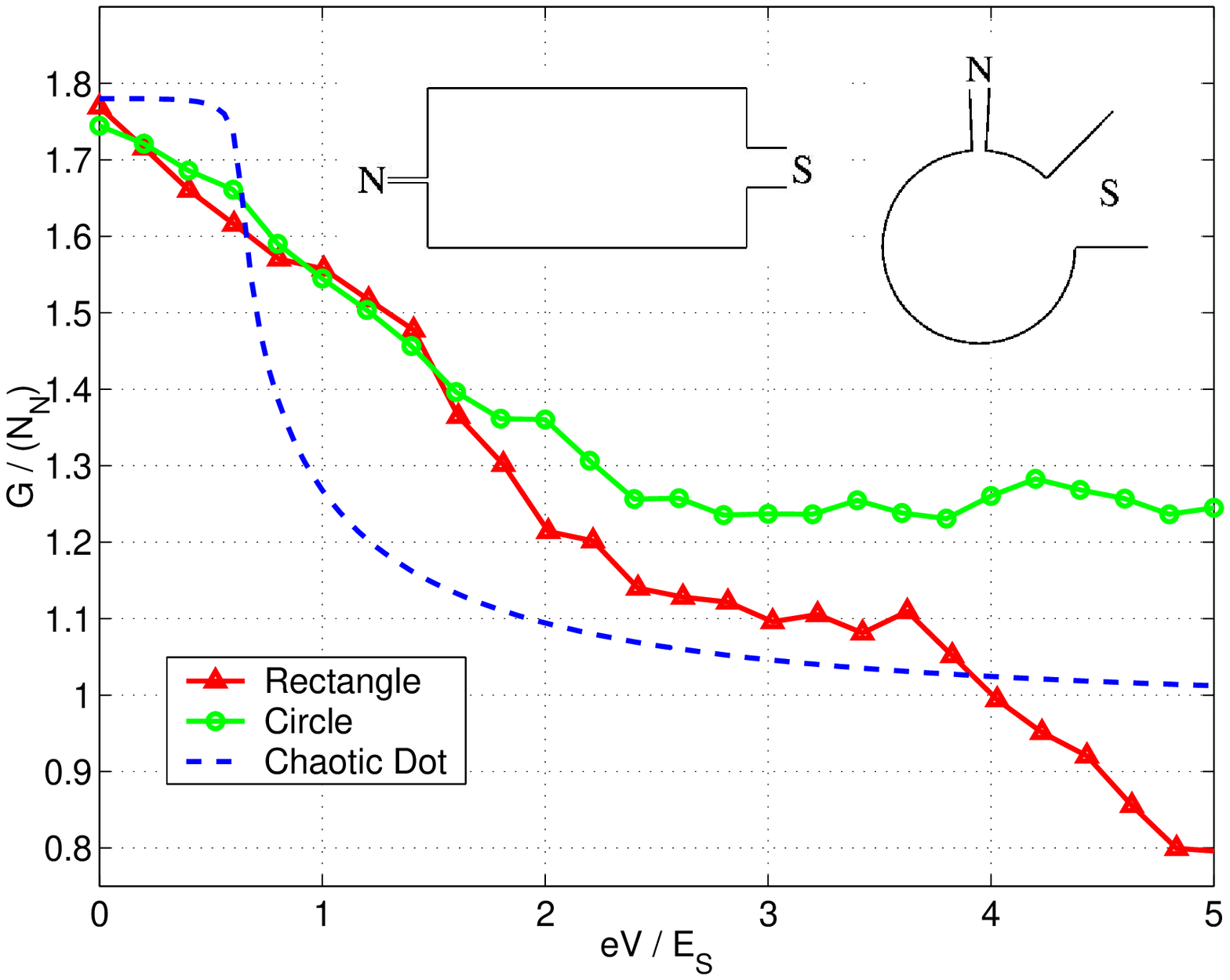,width=8.5cm}}
\refstepcounter{figure} \label{IntComparefig} {\small FIG.
\ref{IntComparefig} Conductance (scaled by $N_N$) vs. voltage
for a rectangular dot and a circular dot, each with $N_N = 10$,
$N_S = 30 N_N$ (shapes shown in inset, with the size of the normal
lead exaggerated).  Each plot was produced by
averaging over small variations in $E_F$. Unlike a chaotic dot
(dashed line), the Andreev conductance of the integrable dots has
no ``plateau'' for $eV<\IGap$. The conductance of the rectangle
drops below $N_N$ due to a weak localization correction.
The energy scale $E_S$ is defined in Eq. (\ref{ESDefn}).}
\end{figure}

In this subsection, we examine the conductance of an N-dot-S
system where the classical dynamics of the dot are integrable.
Previous studies \cite{MELSEN,LODDER} have indicated that the
proximity effect in the density of states is very different for
chaotic and integrable billiards; in the latter, there is no
induced gap, but rather the density of states tends to zero
linearly at $E_F$.  We argue here that the proximity effect's
strong sensitivity to chaos versus integrability also manifests
itself in the Andreev conductance.

We consider two different integrable systems: a rectangular dot
and a circular dot.  In each case, the dot is coupled via narrow
leads to both a normal metal and a superconductor; there are no
tunnel barriers.  As in the previous subsection, we consider the
case where the width of the normal contact is much smaller than
the width of the superconducting contact (see inset of Fig.
\ref{IntComparefig}), so that the normal contact serves
as a probe of the proximity effect in the quantum dot.
The scattering matrix of the system is
computed by numerically matching wavefunctions across the
structure, and the conductance then follows from equation
(\ref{TEFormula}).  The number of modes in the normal lead was
fixed at $N_N = \text{int}\left(k_F W / \pi \right)$
\cite{CIRCNOTE}; we averaged results over small variations in
$k_F$ which did not change $N_N$. Our results are displayed in
Fig. \ref{IntComparefig}; similar results are obtained if one
changes the position of the two point contacts.

The results for the two integrable systems are similar to one
another, and differ significantly from what was found for
a chaotic dot-- as opposed to flat sub-gap region followed by a
rapid drop-off, we have instead a gradual linear reduction of the
conductance with voltage.  We argue that the difference exhibited
here is generic for integrable systems. The reason is the same as
that given in \cite{MELSEN} to explain the difference in density
of states, namely, that the distribution of path lengths is very
different for integrable and chaotic dynamics \cite{Bauer}.  In
the first case, there is a power law distribution of path lengths,
meaning that there is an appreciable probability to find long
paths.  Even for small voltages, these long paths will quickly
acquire a random phase, leading to destructive interference and a
reduction of the conductance. In contrast, chaotic systems have an
exponential distribution of path lengths-- long paths are quite
rare.  In this case, small voltages will not be able to cause any
significant phase-randomization, and thus there will be no
resulting destructive interference of Andreev trajectories.

Though the semiclassical reasoning in terms of path lengths
presented here provides a qualitative account for the difference
between the chaotic and integrable Andreev conductance lineshapes,
attempts to translate it into a quantitative theory have not been
successful.  As was demonstrated in the previous section, a simple
semiclassical theory for the chaotic case fails to recover the
correct lineshape.  A similar problem is encountered when one
tries to do the analogous calculation for the integrable case--
such a calculation predicts the Andreev conductance should fall
off quadratically with voltage, albeit at a faster rate than 
in the chaotic case.  This behaviour is clearly at odds with our
exact calculation showing a linear dependence on voltage.
In many ways, the failure of a semiclassical approach is not
surprising. It is known that semiclassical approximations are
unreliable in superconducting systems, as the usual diagonal
approximation is worse than for normal systems \cite{MELSEN,LODDER}.

Finally, we note that there is a striking connection between the
Andreev conductance's sensitivity to the nature of the dot's
classical dynamics and weak localization.  It has been shown both
theoretically \cite{BARANGER} and experimentally \cite{CHANG} that
the magnetic field dependence of the weak localization correction
of a quantum dot with two normal leads is very different for
chaotic versus integrable dots.  In the chaotic case, one finds a
smooth Lorentzian field dependence, while in the integrable
case a much sharper profile is found, with a cusp at zero
magnetic field.  The similarity between the Andreev conductance
effect and that in weak localization is not coincidental; both
effects rely on the interference of time reversed paths.

Despite this strong similarity, it is worth noting that the effect
in the Andreev conductance is much more pronounced. Here, the
difference between the chaotic and integrable lineshapes is more
severe than in the weak localization effect (i.e. the chaotic case
is much flatter than a Lorentzian). The magnitude of the effect is
also much larger in the Andreev case; the signature of chaos
versus integrability is the {\it entire conductance} itself, not a
quantum correction like weak localization.  For this reason,
the effect in the Andreev conductance should be observable in a
single sample, whereas an ensemble average is required in the weak
localization case.

\section{Re-entrance Effects}
We shift focus in this section, and examine so-called
``re-entrance'' phenomena in the Andreev conductance of a chaotic quantum
dot.  These effects are loosely defined by non-monotonic behaviour
of the conductance in either voltage or magnetic field.  They are
well known in the case of diffusive NS systems
\cite{NAZAROV,CHARLAT}, where the word ``re-entrant'' is used
because the Andreev conductance is the same as the normal
conductance at zero voltage and magnetic field, and at
high voltage or field, but {\it not} for intermediate values.  The
theory developed here allows us to address this behaviour using
a scattering approach, whereas previous approaches almost
exclusively made use of the quasi-classical Green function
technique.  In what follows, we discuss the cases of ballistic
contacts ($T_N = T_S = 1$) and tunnel contacts ($T_N,T_S \ll 1$)
separately.

\subsection{Ballistic Contacts}
In the absence of tunnel junctions, the equations determining the
conductance simplify considerably. We find:
\begin{equation}\label{BALLdrude}
  g_{\rm Dir}(\ve) = 2 N_N
  \tan^2\left(\frac{\theta(\ve)}{2}\right),
\end{equation}
\begin{eqnarray}\label{BALLdiff}
  g_{\rm Diff}(\ve) & = & \frac{N_N^2}{2} \left| \cos\left(\frac{\theta(\ve)}{2}\right)
  \right|^{-4} \times   \\
  & &  \frac{
        N_N \left|\tan\left(\frac{\theta(\ve)}{2}\right)\right|^2 +
            N_S \left|\frac{\cos(\theta(\ve))}{1+\sin(\theta(\ve))} \right|^2}
        { \Lambda(\ve)^2 - \Omega(\ve)^2 }.
        \nonumber
\end{eqnarray}

As discussed in the previous section, these two contributions to
the Andreev conductance can be interpreted as representing two
distinct physical consequences of the proximity effect.  The
direct term $g_{\rm Dir}$ represents processes in which the dot
mimics a bulk superconductor, with Andreev reflections
effectively occurring locally at the N-dot interface. It {\it
decreases monotonically} with voltage and magnetic field, going to
zero at large voltages or magnetic fields. This reflects the fact
that the induced superconductivity effect is sensitive to the
averaged $\it amplitude$ for Andreev reflection, which is largest
at $V=B=0$. On the other hand, the diffusion term $g_{\rm Diff}$ {\it
increases} monotonically with $V$ and $B$, tending to the
classical result for two conductances in series:
\begin{equation}\label{GClass}
  g = g_{\rm Diff} = \frac{(2 N_S)(N_N)}{2N_S + N_N}.
\end{equation}
It represents a contribution from Andreev quasiparticles in the
dot, and is thus sensitive to the dot's density of states.  In the
present case of ballistic contacts, the density of states does not
have any BCS-type peak and thus the diffusion term rises steadily
with applied $V$ or $B$.

Given that the direct and diffusion contributions react in
opposite fashions to an increase in $V$ or $B$, it is not
surprising that a non-monotonic $V$ or $B$ dependence of the total
conductance can be found if the relative strengths of these two
terms are varied.  The latter can be achieved by tuning the ratio
of $N_N / N_S$.  If $N_N \ll N_S$, the direct term will be dominant
at $V,B=0$, and we expect a monotonic decrease of $G$ as a
magnetic field or finite bias are applied.  In the opposite limit
$N_N \gg N_S$, the diffusion term dominates at $V,B=0$, and $G$ is
expected to increase with $V$ or $B$.  A non-monotonic $V$ or $B$
dependence can thus be anticipated in the intermediate regime
where $N_N$ and $N_S$ are comparable.

To quantify the competition between the direct and diffusion
contributions, we examine these terms at
 $V=B=0$.   This is done by solving the
self-energy equations (\ref{ASE1})-(\ref{ASE5}) to determine
$\theta(\ve)$, and then substitute this into Eqs.
(\ref{BALLdrude}) and (\ref{BALLdiff}). Letting $N_{\rm Sum} =
\sqrt{N_N^2 + 6 N_N N_S + N_S^2}$, we obtain:
\begin{mathletters}
\begin{eqnarray}
    g_{\rm Dir} & = & (N_{\rm Sum} - N_S - N_N) \frac{N_{\rm Sum}}{N_N} - 2N_S
\\
    g_{\rm Diff}
& = & (N_{\rm Sum} - N_S - 3 N_N) \frac{N_S}{N_N} + \frac{4 N_N
N_S}{N_{\rm Sum}}
\end{eqnarray}
\end{mathletters}%
with the total conductance given by \cite{BROUWERJMP}
\begin{equation}\label{BTotalG}
    g  =  g_{\rm Dir} + g_{\rm Diff} =
(N_S + N_N) \left(1 - \frac{N_N  + N_S}{N_{\rm Sum}}\right)
\end{equation}

\begin{figure}[t]
\centerline{\psfig{figure=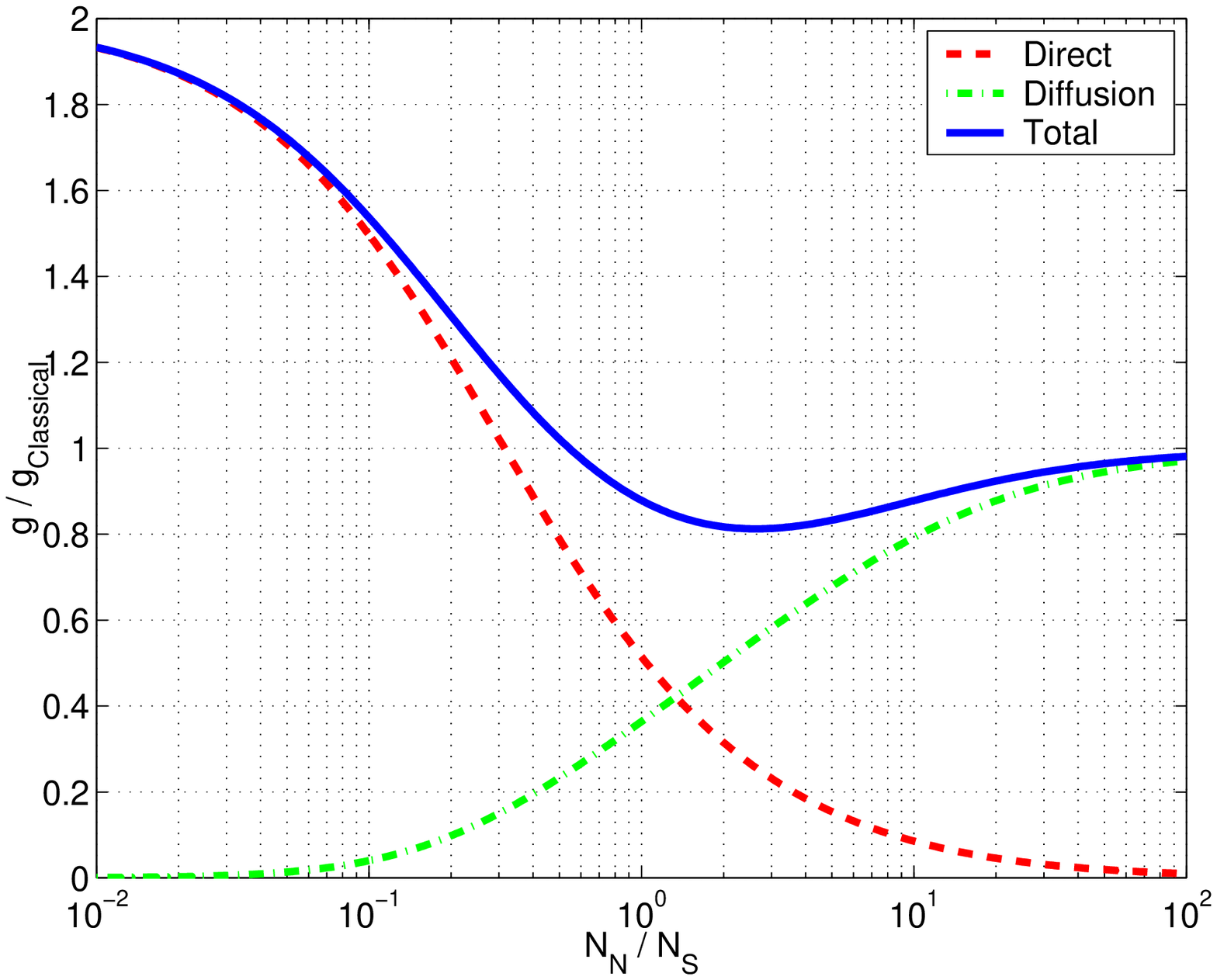,width=7.5cm}}
\refstepcounter{figure} \label{TermsPlot} {\small FIG.
\ref{TermsPlot}. Plot of the direct and diffusion contributions to
the $V=0,B=0$ conductance as a function of $N_N/N_S$ for
transparent point contacts ($T_N=T_S=1$).  At $N_N \simeq
\frac{1}{2} N_N$, the total $V=0,B=0$ conductance is the same as
the classical result $g_{class}$. }
\end{figure}

\begin{figure}[t]
\centerline{\psfig{figure=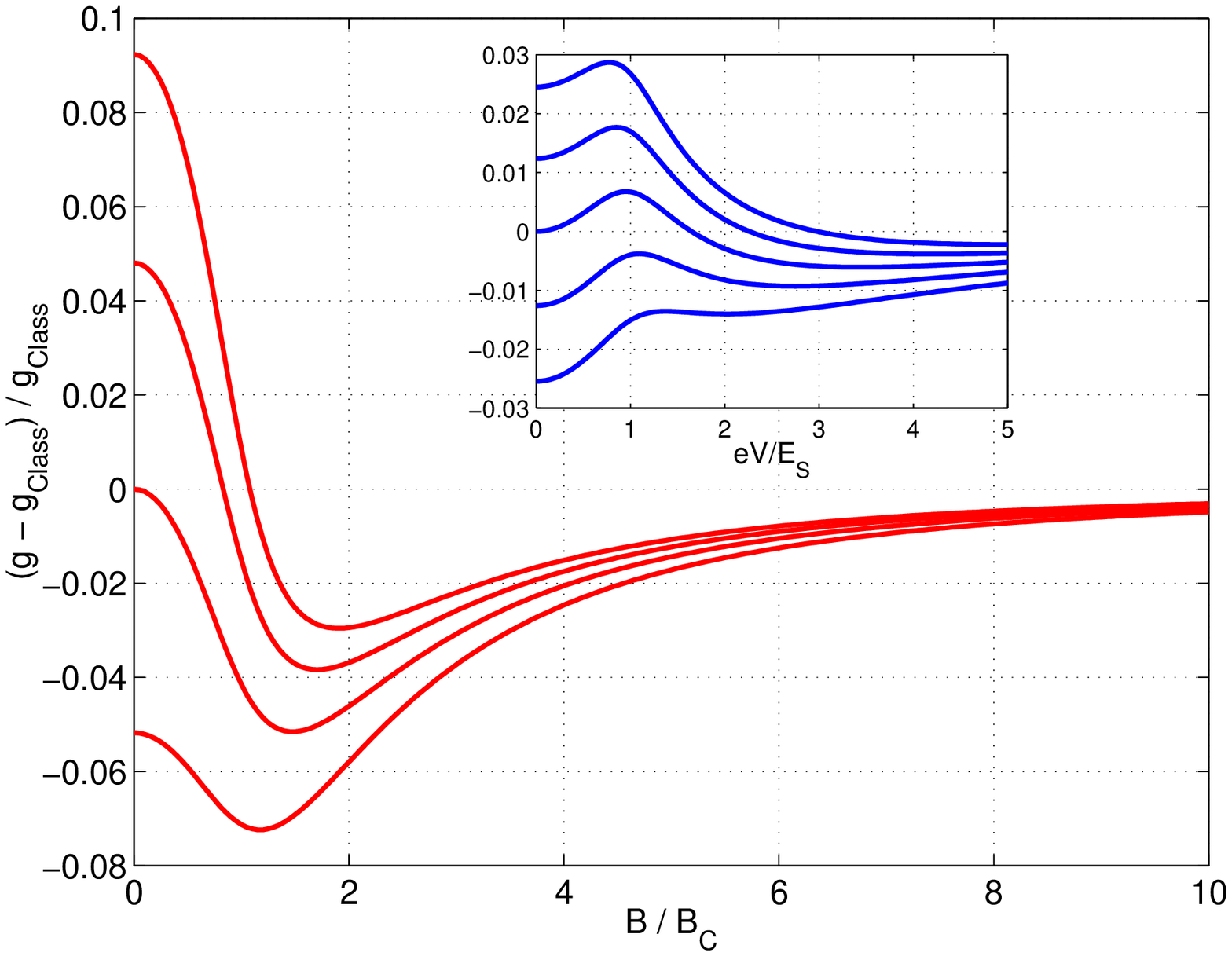,width=8.5cm}}
\refstepcounter{figure} \label{BallReentfig} {\small FIG.
\ref{BallReentfig}. Plot of the excess conductance
$(g(B,V)-g_{class})/g_{class}$ as a function of $B$ at $V=0$,
showing non-monotonic behaviour. From top to bottom, curves are
for $N_N / N_S = 0.39,0.45, 0.54, 0.68$. The inset shows the
excess conductance at $B=0$ as a function of voltage; from top
to bottom, curves are for $N_N / N_S = 0.49, 0.51,0.57, 0.6$.  In
general, re-entrance effects are maximized at $N_N / N_S \simeq
0.54$, where the $V=0,B=0$ conductance is the same as
that large $V$ or $B$.
}
\end{figure}

Plotted in Figure \ref{TermsPlot} as a function of $N_N/N_S$ are
the zero field values of $g_{\rm Dir}$ , $g_{\rm Diff}$ and $g$,
normalized by the high field conductance given by the
classical formula (\ref{GClass}).  As $N_N/N_S$ is increased from
zero, the total conductance initially decreases with $g_{\rm Dir}$
as the local Andreev reflection effect is suppressed, while at
larger values it starts to increase with $g_{\rm Diff}$ as the
density of states at $E_F$ returns to its normal state value.  At
$N_N \simeq \frac{1}{2} N_S$, we find that the total
conductance at zero field is the same as the classical result. Not
surprisingly, we find that re-entrance effects in both magnetic
field and voltage are maximized here (see Figure
\ref{BallReentfig}).

It is instructive to make a comparison at this point to the
non-monotonic re-entrance behaviour of diffusive NS systems (i.e.
a diffusive normal metal in good contact with a superconductor).
In the diffusive case, the $0$ field and high field conductances are the
same, being equal to the normal state conductance.  The lack of
any change due to superconductivity at $V=B=0$ is usually
explained as an exact cancellation of the conductance doubling
effect of Andreev reflection by a suppression of the density of
states at the Fermi energy.  A conductance enhancement is however
found at finite voltage, roughly at $eV \simeq E_T = \hbar D /
L^2$, where $D$ is the diffusion constant and $L$ is the length of
the normal metal. Quasi-classical calculations find maximum
conductance enhancements on the order of $10\%$ of the normal
state conductance \cite{NAZAROV}.

In the present system, we can tune the relative significance of
the Andreev reflection enhancement and density of states
suppression terms by varying $N_N / N_S$. We find that
non-monotonic effects are maximized when we adjust this ratio to
mimic the diffusive system, by insisting that the $0$ field and
high field conductances are the same; this occurs at $N_N
\simeq 1/2 N_S$. The conductance maximum in voltage is smaller
however, being on the order of $0.01 G_{Classical}$, and occurs
roughly at $eV = E_S = N_S \delta / 2 \pi$.  This is the inverse
of the time needed to escape to the superconductor, and is thus
the analog of $E_T$ in the diffusive system, which represents the
inverse of the time needed for a particle to diffuse across the
normal metal and reach the superconductor.  Note also that if the
normal lead was removed, the size of the induced gap $\IGap$ in
the dot density of states is $\sim E_S$.

Finally, we also find pronounced non-monotonic behaviour in
magnetic field for this range of $N_N/N_S$, with the magnitude of
the effect being larger than that in voltage (Fig. \ref{BallReentfig}).
The maximum
conductance occurs roughly at a flux $\Phi_C$ given by Eq.
(\ref{critflux}), the same flux that would be required to close
the gap in the density of states in the absence of the normal
lead.

\subsection{Tunnel Regime}
We turn now to the case where both point contacts contain opaque
tunnel barriers ($T_S,T_N \ll 1$).  In this regime, it is
sufficient to consider the conductance to lowest non-vanishing
order in $T_N$ and $T_S$.  Of particular interest here is the
well-known ``reflectionless tunneling'' effect \cite{vanWees}-- the
$V=0,B=0$ Andreev conductance of a N-I-N-I-S structure (where $I$
is an insulating region having transmission $T$) is found to be
proportional to $T$, not to $T^2$ as one has for a single barrier.
It is as though the Andreev reflected hole is not reflected at all
by the tunnel barriers.  A similar effect occurs for a N-I-S
system where the normal region is sufficiently disordered. The
origin of this striking behaviour is now understood to result from
the constructive interference of trajectories which reflect many
times between the two barriers, leading to a fraction $\sim T$ of
the conductance channels being open (i.e. having a transmission
probability close to unity) \cite{MELSENBLAH}.

In the present case, we are able to examine the effects of finite
voltage and magnetic field on reflectionless tunneling when a
quantum dot separates the barriers.  At large voltages or magnetic
fields, the pairing angle $\theta(\ve)$ tends to its normal-state value
of 0, and we find that the conductance is given to leading order
by the classical series addition formula:
\begin{equation}\label{TGClass}
  g_{\rm class} = \frac{ (\frac{1}{2} N_S T_S^2) (N_N T_N)  }{\frac{1}{2} N_S T_S^2 + N_N T_N}
\end{equation}
For $N_S T_S^2 \ll N_N T_N$, this simplifies to $g_{\rm class} =
\frac{1}{2} N_S T_S^2$.  Unlike the case at $V=0$,$B=0$, there is
no order $T$ reflectionless tunneling contribution here, as the
necessary constructive interference is lost when electron-hole
degeneracy or time-reversal symmetry is broken. Below we show that
reflectionless tunneling does survive at small values of $B$ and
$V$, and describe how this contribution evolves as $B$ and $V$ are
increased.

We begin by solving Eqs. (\ref{ASE3})-(\ref{ASE5}) for the self
energy to lowest order in $T_N,T_S$ for $B=0$; we find the pairing
angle is given simply by:
\begin{equation}\label{TTheta}
  \theta(\ve) = \arctan\left(\frac{E_S}{E_N - i \ve}\right).
\end{equation}
$E_S$ and $E_N$ are the inverse escape times to the superconductor
and normal metal leads respectively, and are defined by:
\begin{equation}\label{EDefn}
  E_N = \frac{N_N T_N \delta}{2 \pi},  E_S = \frac{N_S T_S \delta}{2
  \pi}.
\end{equation}
Using this result for $\theta(\ve)$, we next write Eqs.
(\ref{gDrude})-(\ref{gD1}) for the conductance to lowest order in
$T_S,T_N$.  The direct contribution $g_{\rm Dir}$ corresponds to
Andreev reflection at the $N$-dot interface and is order $T_N^2$.
The only order $T$ contribution is found in the diffusion term
$g_{\rm Diff}$, which yields:
\begin{equation}\label{TCond}
  g_{\rm Diff}(\ve) = N_N T_N  \frac{E_N E_S^2}{\ETild^2(\ve)}
    \sqrt{ \frac{2}{\ETild^2(\ve) + E_N^2 + E_S^2 - \ve^2  }  } + O(T^2),
\end{equation}
where we have defined:
\begin{equation}\label{ETild}
  \ETild(\ve) = \left[ \left(E_S^2 - E_N^2 - \ve^2\right)^2 + 4\left(E_S
  E_N\right)^2 \right]^\frac{1}{4}.
\end{equation}

\begin{figure}[t]
\centerline{\psfig{figure=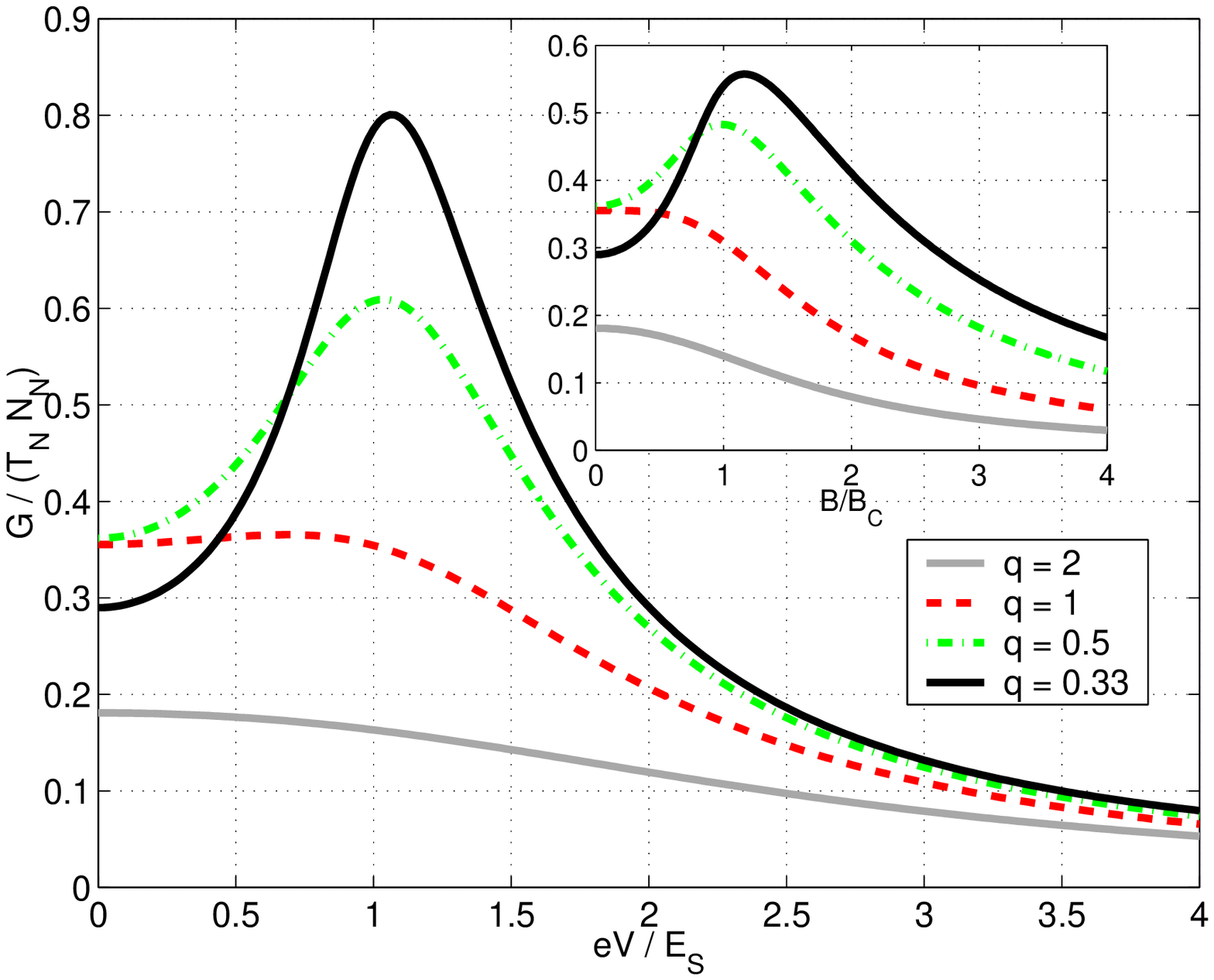,width=7.5cm}}
\refstepcounter{figure}
\label{TVfig}
{\small FIG. \ref{TVfig}.
Conductance vs. voltage in the tunnel regime, $T_N = T_S = 0.01$,
exhibiting the enhancement of reflectionless tunneling at finite voltage.
The parameter $q$ is defined as the ratio $N_N / N_S$.
The inset shows the $V=0$ conductance as a function of magnetic
field for the same parameter choices.}
\end{figure}

Eq. (\ref{TCond}) gives the complete voltage dependence of the
reflectionless tunneling effect.  At $V=0$, it reduces to:
\begin{equation}\label{RT0V}
  g_{\rm Diff}(0) = \frac{ \left(N_N T_N\right)^2 \left(N_S T_S\right)^2
   }{ \left[  \left(N_N T_N\right)^2  +  \left(N_S T_S\right)^2  \right]^{\frac{3}{2}}},
\end{equation}
which is similar to the formula found in \cite{MELSENBLAH},
generalized to the case where the point contacts have different
widths.  Depending on the relative magnitudes of $E_S$ and $E_N$,
the conductance drops monotonically with voltage, or shows a
maximum around $eV \simeq E_S$ (see Fig. \ref{TVfig}).  The
location of the maximum is at $eV = \sqrt{7/6} E_S$ if $E_S \gg
E_N$. Similar behaviour is found for the magnetic field dependence
of the conductance (see inset of Fig. \ref{TVfig}).

It is easy to understand the origin of this non-monotonic
behaviour within our theory.  When $N_S$ becomes larger than
$N_N$, the effect of the superconductor on the dot density of
states becomes significant. As we have seen in Section II, the
induced density of states will have a sharp peak at $E_S$ when
$T_S \ll 1$; it is this peak that is manifesting itself in the
conductance.

It is interesting to note that Eq. (\ref{TCond}) has the same form
as what was found for a diffusive N-I-N-I-S system using a
quasi-classical Green's function approach \cite{Volkov}.  In that
system, the quantum dot is replaced by a diffusive normal wire of
length $d$, width $W$ and mean free path $l$, and the normal and
superconducting leads are not attached via point contacts, but are
also wires of the same width. In the limit $T_N,T_S \ll l/d$ the
resistances of the tunnel barriers dominate, and an expression
identical to Eq. (\ref{TCond}) was obtained, but now the energies
$E_N,E_S$ are given by:
\begin{equation}\label{DiffEnerg}
  E_N = T_N \frac{\hbar v_F}{4 d}  ,  E_S = T_S \frac{\hbar v_F}{4 d}
\end{equation}
where $v_F$ is the Fermi velocity in the normal inter-barrier
region.  This is identical to the definition of $E_N$ and $E_S$ in
Eq. (\ref{EDefn}) if one takes $\delta$ to be the level spacing in
the wire.  Note that this correspondence is not surprising;
previous studies have also found that adding strong tunnel
barriers makes a diffusive normal wire with many channels
equivalent to a quantum dot.

\section{Conclusions}
We have studied the voltage ($V$) dependence of the Andreev
conductance of chaotic and integrable quantum dots; we have also
examined the magnetic field ($B$) dependence in the chaotic case.
In the regime where the contact to the superconductor dominates,
we find that the voltage dependence of the Andreev conductance is
extremely sensitive to the nature of the dot's classical
dynamics-- in the chaotic case, the dot itself mimics a
superconductor and the conductance is initially flat in voltage,
whereas in the integrable case the conductance falls off linearly
with voltage. This effect is large in that the entire conductance
forms the signature of chaos vs. integrability; also, it does not
require any ensemble averaging to be done.  Both these facts make
it particularly amenable to experiment.  Also important in this
regard is that one does not require an extremely clean contact
between the dot and the superconductor as long as the contact is
large.

We have also studied non-monotonic ``re-entrance'' phenomenon in the
$V$ and $B$ dependence of the Andreev conductance of a chaotic
dot. We find that such behaviour is ubiquitous, and is the result
of two competing processes-- Andreev reflection at the dot-normal
lead interface, which decreases as $V$ or $B$ increases, versus
quasiparticles being injected into the dot, which increases with
$V$ and $B$.

\section{Acknowledgements}

We thank C. W. J. Beenakker for a useful discussion.
A.C. acknowledges the support of the Olin Foundation and the
Cornell Center for Materials Research. Work supported in part by
the NSF under grant DMR-9805613.

\appendix
\section{Calculating the Andreev Conductance for $B \ne 0$}

In this appendix, we outline the method used for calculations at
non-zero magnetic field $B$.  Formally, the magnetic field
enters the model in a different fashion than a voltage difference.
The latter is dealt with by the fact that the Andreev scattering matrix
$\SS$ defined in Eq. (\ref{SSDefn}) is an explicit function of energy.
The field dependence however does not appear directly in the
expression for $\SS$, but rather only emerges in the averaging procedure--
as we use the Pandey-Mehta distribution defined in Eq. (\ref{PandeyDist}),
the ensemble of random matrices is itself a function of field.

As in the calculation at $B=0$, the first step in obtaining the
conductance is to calculate the averaged matrix Green function
$\langle \GG \rangle$ defined in Eq. (\ref{GDyson}).  The diagrams
used are the same as the $B=0$ case, but now use of the
distribution (\ref{PandeyDist}) leads to a different self-energy
$\Sigma$ in the Dyson equation:

\begin{eqnarray}
    \label{PMAvgSE}
        \Sigma(\ve,\gamma) & = & 1_M \otimes \left( \begin{array}{cc}
                            \Sigma_{ee} & (1-2\gamma) \Sigma_{eh} \\
                            (1-2\gamma) \Sigma_{he} & \Sigma_{hh}
                        \end{array} \right)          \\
        \nonumber
               & = &  1_M \otimes \frac{\lambda^2}{M}
                            \left( \begin{array}{cc}
                            \langle \tr \GG_{ee}\rangle & (2\gamma-1) \langle \tr \GG_{eh}\rangle \\
                          (2\gamma-1) \langle \tr \GG_{he}\rangle & \langle \tr \GG_{hh}\rangle
                        \end{array} \right),
\end{eqnarray}
where $\gamma$ is a function of magnetic field (see Eq.
\ref{FluxReln}).  The additional factors of $(1-2\gamma)$ here
reflect the fact that breaking time-reversal symmetry suppresses
off-diagonal superconducting correlations.  Solving the Dyson
equation (\ref{GDyson}) to leading order in $1/M$, we find that
relations (\ref{ASE1}) and (\ref{ASE2}) continue to hold, meaning
that we may still parameterize the self-energies in terms of a
pairing angle $\theta(\ve,\gamma)$ through Eq. (\ref{trig}). The
equation determining $\theta(\ve,\gamma)$ now takes the form:
\begin{equation}\label{PMTheta}
    \tan(\theta(\ve,\gamma)) =  \frac{N_S Q_S - 2 \gamma \sin(\theta)}
{N_N Q_N -i \frac{\pi \ve}{2 \delta} },
\end{equation}
where $Q_N$ and $Q_S$ are functions of $\theta$ defined in Eqs.
(\ref{ASE4}) and (\ref{ASE5}).  The $1/M$ corrections to the
self-energy in the presence of a magnetic field read: \bleq
\ifpreprintsty\else \renewcommand{\thesection}{\Alph{section}} %
\renewcommand{\theequation}{\Alph{section}\arabic{equation}} \fi %
\begin{eqnarray}\label{BSENorm}
\frac{\Seh^2 - \See^2}{\lambda^2} - 1   =
   -\frac{1}{M} \bigg(N_S Q_S \left(\sin(\theta) + Z_S \right)
 \mbox{} + N_N Q_N \left(\cos(\theta) + Z_N \right)
          + i \frac{\pi \ve}{2 \delta} + 2 \gamma \sin^2(\theta)
                    \bigg).
\end{eqnarray}
\eleq

The next step in the calculation is to sum diagrams for the
conductance. The necessary diagrams are the same as those retained
in the $B=0$ calculation (i.e. direct and diffusion terms),
although their evaluation is different.  The direct term is still
given by Eq. (\ref{gDrude}), but with $\theta(\ve,\gamma)$ now
determined from Eq. (\ref{PMTheta}).

The diffusion term acquires a form different from Eq. (\ref{gD1}),
as factors of $(1-2\gamma)$ now appear in graphs where the
particle-hole indices of upper and lower branches do not match.
These factors lead to $1/M$ corrections both to the matrix inverse
that arises when summing the diffusion ladder, and to the matrix
prefactors of the ladder.  As discussed earlier, such corrections
are important when calculating the conductance.  The result is:
\bleq

\begin{eqnarray}
 \label{Bgdiff}
g_{\rm Diff} = 8 N_N^2 \left| \frac{Q_N^2}{Z_N} \right|^2 \frac{
\left(1 - Z_N^2 \right)^2 \tilde{\Lambda} | \sin(\theta) |^2
       + (\Pi_N + \gamma \Pi_B) |\sin(\theta)|^2 + \Pi_S |\cos(\theta)|^2 }{\tilde{\Lambda}^2 -
    \tilde{\Omega}^2},
\end{eqnarray}
where
\begin{eqnarray}
\label{BNumCorr}
\Pi_B(\theta)  & = & 2 \Big((1+Z_N)^2 \left( 1 + 2 \text{Re}
[\sin^2(\theta)] \right) + 4 Z_N (1 + Z_N) \text{Re}[\cos(\theta)]
+ 4 Z_N^2 \left( \left| \cos(\theta) \right|^2 - 2\text{Re}[\sin^2(\theta)] \right)
        \Big),  \\
\label{BLambdaCorr}
    \widetilde{\Lambda}(\theta,\gamma) & = & \Lambda(\theta) - 12 \gamma \sin^2(\theta),  \\
\label{BOmegaCorr}
    \widetilde{\Omega}(\theta,\gamma) & = & \Omega(\theta)
    + 4 \gamma \left( 1- \left| \cos(\theta) \right|^2 + 4\text{Re}[\sin^2(\theta)] \right)
    \widetilde{\Lambda}(\theta,\gamma)
    + 8 \gamma \left(
        N_N \left| \frac{Q_N}{\GN} \right|^2 Y_N(\theta)
        + N_S \left| \frac{Q_S}{\GS} \right|^2 Y_S(\theta)
        + \gamma Y_2(\theta)
        \right),
 \\
\nonumber
Y_N(\theta) & = &
  \left( \left| \frac{Z_N}{Q_N} \right|^2 - 1 \right)
    \left( 1- \left| \cos(\theta) \right|^2 + 4\text{Re}[\sin^2(\theta)] \right)
- 2 Z_N \Big( 5 \text{Re}[\sin^2(\theta)]\text{Re}[\cos(\theta)] +
\text{Im}[\sin^2(\theta)]\text{Im}[\cos(\theta)]  \Big)\\
\label{YN}
 &&
  + Z_N^2 \Big( 1 - \left|\cos(\theta)\right|^2
         -2 \text{Re}[\sin^2(\theta)]\left(1 + 2\left|\cos(\theta)\right|^2 \right) \Big), \\
Y_S(\theta) & = & \left(
            \left| \frac{Z_N}{Q_N} \right|^2 - \left| 1 - Z_S \sin(\theta)\right|^2
          \right)
          \left( 1- \left| \cos(\theta) \right|^2 + 4\text{Re}[\sin^2(\theta)] \right) -
        Z_S^2 \left|\sin(\theta) \cos(\theta) \right|^2,  \\
\label{Y2}
Y_2(\theta) & = & 1
  - \left|\cos(\theta)\right|^2 \left(1 + 4 \text{Re}\left(\sin(\theta)^2\right) \right)
  + \text{Re}\left[ 3\sin^2(\theta) + 4\sin^4(\theta) \right]
  + 4 \left| \sin(\theta) \right|^4.
\end{eqnarray}
\eleq Here $\Lambda(\theta)$, $\Omega(\theta)$, $\Pi_N(\theta)$
and $\Pi_S(\theta)$ are given by Eqs. (\ref{LambdaKernel}) -
(\ref{PiSKernel}).  The above equations, together with Eq.
(\ref{PMTheta}) for $\theta(\ve,\gamma)$, determine the Andreev
conductance for arbitrary voltage and magnetic field.

\ecols


\begin{references}
\bibitem{DECOMPNOTE}
A similar decomposition can be made for transport through a
normal dot.  In this case, the first term corresponds to
direct reflection off tunnel barriers in the contacts,
while the second contribution is for processes that
explore the dot ergodically.  See i.e. \cite{BeenRMT}.
\bibitem{CIRCNOTE}
Note that in the case of a circular dot, we make use of
wedge-shaped leads to preserve the symmetry of the problem.  In
this case, the number of propagating modes in the leads depends on
the distance from the center of the circle.  We may truncate the
scattering modes because higher order modes (with classical
turning points outside the billiard) do not couple to the dot, and
thus do not contribute to the Andreev conductance.
\end{references}
\end{document}